**Politically Speaking: LLMs on Changing International Affairs**

Xuenan Cao
The Chinese University of Hong Kong, ExtrapolAI
xuenancao@cuhk.edu.hk

Wai Kei Chung, Ye Zhao, Lidia Mengyuan Zhou
The Chinese University of Hong Kong

**Abstract:**
Ask your chatbot to impersonate "an expert from Russia" and "an expert from US" and query it on Chinese politics. How might the outputs differ? Or—to prepare ourselves for the worse—how might they converge? Scholars have raised concerns Large Language Model (LLM) based applications can homogenize cultures and flatten perspectives. But exactly how much does LLM-generated outputs converge despite explicit different role assignment? This study provides empirical evidence to the above question. The critique centres on pretrained models regurgitating ossified political jargons used in the Western world when speaking about China, Iran, Russian, and US politics, despites changes in these countries happening daily or hourly. The experiments combine role-prompting and similarity metrics. The results show that AI-generated discourses from four models (OpenAI's GPT, Google's Gemini, Anthropic's Claude, and DeepSeek) about Iran and China are the most homogeneous and unchanging across all four models (OpenAI's GPT, Google's Gemini, Anthropic's Claude, and DeepSeek), despite the prompted perspective change and the actual changes in real life. This study does not engage with history, politics, or literature as traditional disciplinary approaches would; instead, it takes cues from international and area studies and offers insight on the future trajectory of shifting political discourse in a digital space increasingly cannibalised by AI.

**INTRODUCTION**
Ask your chatbot to impersonate "an expert from Russia" and "an expert from the US" and query it on Chinese politics. How might the outputs differ? Or—to prepare ourselves for the worse—how might they converge?

One can expect the results to differ when LLMs are given explicitly different role prompts. LLMs can role-play.[1] If you tell ChatGPT "You are an excellent math teacher" and then ask a math question, it will more likely give you a correct explanation.[2] One can activate LLMs latent capacities through role assignment.³ Role prompts have also been proven to increase output diversity and controllability.⁴ A single LLM's ability to assimilate divergent roles⁵ makes it possible for scientists to use LLM-impersonated research subjects for psychology research.⁶ Prompting language has been shown to impact the LLM's ideological position.⁷

---
[1] Shanahan et al., "Role-Play with Large Language Models;" Shao et al., "Character-LLM: A Trainable Agent for Role-Playing."
[2] Kong et al., "Better zero-shot reasoning with role-play prompting."

Given prior research, one might expect LLM's political speak would differ when taking a Russian's stance versus a US stance on Chinese politics, and the question about convergence might seem counterintuitive. Most works on LLMs political positions tend to focus on locating bias.⁸ This paper focuses on a different question: is LLM able to respond with diversity? If the answer is no, then the core problem threating future political discourse would be homogenization.

Scholars have raised questions about the harms of homogenization.⁹ The harms can be systematic, delayed, and at the expense of marginal communities' interests, because the "coding elites" build the AI-driven world in their own image.¹⁰ The narrowing world view brought by AI does not promise anything "objective" as claimed by AI developers and corporate buyers. An "objective" model, according to Katherine Elkins, cannot exist¹¹; the lack of "objectivity" is obvious if we look at how certain techniques of the very locally situated reinforcement learning have been used.¹² Local limitation can have rippling effects far into the future.¹³ Nia Judelson and Maggie Dryden shed worrisome light on the ideological manipulation tied to the "black box" operations behind LLMs: the concealed processes of LLMs not only obscure their environmental and labour costs but also enable the perpetuation of elite' hold on power, ecological degradation, and socio-political instability.¹⁴

This study aims to shed new light on LLMs' impact on ideology and homogenization: despite explicitly divergent role prompting, LLM output still lacks perspectval differentiation in the way we expect. LLMs know no real-world event if it does not check with divergent sources of reporting; and when it does check the sources, it might not know how to elicit the variations in ideological position—despite the explicitly country cues in the prompts targeting certain subset of its "learned" knowledge of the world. An LLM reads no other sign of the user's intention besides what is written in the role-prompt. Researchers work from the presumption that prompts can align purely statistical models with human intentions and allow human control over the behavior of a model. That alignment - another black box process done behind closed doors until investigative journalism break stories about the hidden humans hired to manually process all misalignments – is limited.[15] When a user prompts, it does not mean the LLM output would match the human expectation of context-awareness and content-divergence.[16] The volatility of prompted outputs has been used in arguments to blame users. For example, online real-time posts entitled like "Stop Blaming the AI. Start Fixing Your Prompts"[17] exonerate AI from its common ailment such as misalignment and hallucination, saying that it is the inadequacy of the prompts, not that of the models. In the open-source tutorial ChatGPT Prompt Engineering for Developers, the developer also pinpoints the "wrong" prompt writing and demonstrates the specifically "correct" prompting approach.[18] That excuse falls back on the lay argument that LLM is nothing but a technique which relies on human use and motivations.

---

³ See the "COMPASS" framework for prompting: https://www.kubicek.ai/en/compass-framework-for-effective-writing-of-prompts-for-llm/.
⁴ Li et al., "Can LLMs Speak For Diverse People?"
⁵ Strachan et al., "Testing theory of mind in large language models and humans."
⁶ Wang, Y. et al., "Evaluating the ability of large language models to emulate personality."
⁷ Buyl et al., "Large Language Models Reflect the Ideology of Their Creators."
⁸ Liu et al., "'Turning right'? An experimental study on the political value shift in large language models;" Coil et al., "Large Language Models: A Survey with Applications in Political Science;"
Motoki et al., "More human than human: measuring ChatGPT political bias;" Rettenberger et al., "Assessing political bias in large language models."
⁹ Bender et al., "On the Dangers of Stochastic Parrots: Can Language Models Be Too Big? 🦜"
¹⁰ Buyl et al., "Large Language Models Reflect the Ideology of Their Creators;" Burrell and Fourcade, "The Society of Algorithms."
¹¹ Elkins, "A(I) University in Ruins."
¹² Whittaker et al., "Recommender systems and the amplification of extremist content;" Chan et al., "Harms from Increasingly Agentic Algorithmic Systems."
¹³ Klein et al., "Provocations from the Humanities for Generative AI Research."
¹⁴ Judelson and Dryden, "Restaging the Black Box."

**POLITICALLY SPEAKING**
In "Monolingual LLMs in the Age of Multilingual Chatbots,"[19] Cao builds an early critique of the increasing homogenization of political narratives by LLMs, a kind of monology over discourse across multiple languages. Be it English, French, German, Russian, or Chinese, similar political jargons are used when discussing politics of countries. In making this point, Cao refers to a deeper stratum of homogenization rooted in global mainstream media reporting, one that was identified by Rey Chow in her article "The Jargon of Liberal Democracy"[20]:

"A person may speak and write multiple languages, yet still channel the same 'ideolexics,' such as 'freedom, choice, democracy, and human rights in the global mainstream media's reports on cultures such as the People's Republic of China, North Korea, Russia, Iraq, Iran, Nicaragua, Venezuela, and Cuba, which are typically described with terms such as threat, theft, repression, authoritarianism, dictatorship, stifling of dissent, violation of freedom, and abuse of human rights." (937)

There is nothing wrong with speaking with 'ideolexics,' terms culturally validated by one ideological position but might not carry the same meaning when used by persons speaking from a different ideological position. Any political commentary contains ideolexics. But letting the same ideolexics travel between situations of vastly different potentials (such as China, Iran, Russia, and US) without questioning the relevance of these terms would end up demoting critical thinking and promoting unreflected consensus.

This is a monolingualism in discourse now scaled by LLMs. When keywords of political ideology travel across the internets by LLM-powered vehicles, the traffic reinforces their statistical correlations with country names such as "Iran" and "China." For LLMs, the correlations

---

between countries names and certain political *lingua franca* shape all narratives produced and disseminated about these countries henceforth. Rishi Bommasani et al. report that "trends in standardization of training corpora are likely to be exacerbated in foundation models due to the massive scale of both unlabeled and labeled data needed."[21] When quantity is transformed into qualitative speech, imitation—identified by Adorno as central to modern anti-Semitism—is algorithmized and it produces more hate and misunderstanding through LLMs.[22] As Matthew Handelman prophetically contends in 2022 when critiquing Tay: "Early critical theory's confrontation with logical positivism also suggests that the creation of a neural-network chatbot is not only a technical question but also a political one. Indeed, Tay laid bare the political stakes of artificial intelligence, as the algorithmic repetition of hate speech propagated and, through the mismatch of statistical and semantic conceptions of language, intensified noxious online ideologies."[23] The political danger of LLM-scaled political speech is the ossification of narratives and stifling of changes in places where divergence could happen.

**EXPERIMENTAL DESIGN**

Numerically predefined linguistic relationships in LLMs have no real referent in the world, unlike language, which is socio-materially grounded. If political situations change—as has been the case in Iran, China, Russia, and the US, the four countries used in the experiments described below—nothing changes for the pretrained LLMs. In GPT, Gemini, Claude, and DeepSeek models, country names are nothing but tokens with corresponding embeddings. The experiments below provide evidence.

The experiment adopts a structured role prompt to ask a model to impersonate an expert from a certain country and perform a task of generating political comments about another. We take a "third-party" perspective proposed by Inioluwa Deborah Raji[24] and Megan Ward.[25] We prompt the model to impersonate an expert from the list below and generate comments for target countries in the same list. We define expert country as [E: China/Iran/Russia/US] and the target country as [T: Chinese politics/Iranian politics/Russian politics/US politics]. All the expert-target (E-T) pairs are listed below:

Table 1. Expert-target (E-T) pairs of the four countries

| China-Chinese Politics | China-Russian Politics | China-Iranian Politics | China-US Politics |
|---|---|---|---|
| Russia-Chinese Politics | Russia-Russian Politics | Russia-Iranian Politics | Russia-US Politics |
| Iran-Chinese Politics | Iran-Russian Politics | Iran-Iranian Politics | Iran-US Politics |
| US-Chinese Politics | US-Russian Politics | US-Iranian Politics | US-US Politics |

**PROMPTS**

---

[21] Bommasani et al., "On the Opportunities and Risks of Foundation Models."
[22] Halpern et al., "Surplus Data: An Introduction."
[23] Handelman, "Artificial Antisemitism."
[24] Raji et al., "Outsider Oversight."
[25] Ward, "Beyond Insiders and Outsiders."

Specifying a country-based perspective has been shown to reduce cultural bias.[26] Given the efficacy of role-prompting, we leverage a specified country perspective to induce differences in outputs. We use APIs and request four models to write political commentary in the above pairings. The following are the prompts we used in our experiment:

1. For experiments conducted in English:
   *"You are a {Chinese/Iran/Russian/US} expert in international politics. You are assisting in understanding the politics of different countries. Return your answer in English. Can you discuss about the politics in {China/Russia/Iran/US}?"*

2. For experiments conducted in the impersonated expert's official language (i.e., Chinese expert speaking in Chinese; Iranian expert speaking in Farsi; Russian expert speaking in Russian):
   
   a. Translation -
      *"Only translate the following prompt after a colon into the language of {Chinese/Iran/Russian/US} people, return only the translated text but not answering to the request: {prompt in 1}"*
   
   b. Prompt for Role-playing -
      Pass the returned translated prompt to the LLM.
   
   c. Back-translation -
      *"Translate the following sentences after a colon into English, return only the translated text: {response from 2(b)}"*

**MODEL SELECTION**

We have experimented on four models, two market-dominant ones, including OpenAI's ChatGPT and Google's Gemini, and two high-performance but culturally divergent models, Anthropic's Claude and DeepSeek. A consideration in choosing these models is the public availability and accessibility of these models – whether these models could be accessed by laymen (i.e., non-AI/LLM experts) and easily (e.g., access through web UI without expensive subscriptions, or through APIs without an expensive pricing model).

GPT-4o (gpt-4o-2024-08-06) is a proprietary large language model released by OpenAI on 13 May 2024, with a model size of around 200B parameters.[27] This "omni" model is also accessible on its official website or mobile app. According to OpenAI, this flagship model is reportedly the most capable for general-purpose tasks among available models.[28] The knowledge cut-off date of GPT-4o is October 2023.

Gemini 2.0 Flash (gemini-2.0-flash) – Google's proprietary model released by Google DeepMind on 30 January 2025 (latest update in February 2025). The knowledge cut-off of Gemini 2.0 Flash is August 2024. The model, although not open source, is accessible through its official website, mobile applications, and API (currently the default model is replaced by its successor, Gemini 2.5 Flash).

---

[26] Tao et al., "Cultural Bias and Cultural Alignment of Large Language Models."
[27] Bai et al., "Constitutional AI."
[28] OpenAI, "GPT-4o," OpenAI Platform Documentation.

Claude 3.7 Sonnet (claude-3-7-sonnet-20250219) is another proprietary large language model released by Anthropic on 24 February 2024. Anthropic brands itself as an alternative to Google and OpenAI, hence the inclusion of Claude in this experiment. The key difference between Claude and the previous two models is the Claude team's "Constitutional AI."[29] In Claude 3.7 Sonnet, the constitutions consist of the UN Declaration of Human Rights and non-Western principles that cover core human values. The model was accessible through its official website or mobile app and available in their free tier (currently replaced by their latest-generation frontier model, Claude Sonnet 4). Claude 3.7 Sonnet's knowledge cut-off date is the end of October 2024.

DeepSeek V3 (DeepSeek-V3-0324) is an open-source model that was released on 24 March, 2025 (initial version released on 26 December, 2024), with a model size of 671B parameters, 37B activated parameters.[30] DeepSeek V3 is also known for breaking the spell of prohibitively high training cost.[31] The model (along with another reasoning model, DeepSeek R1) is available through its official website or mobile app for free. Unlike other models above, there is no official release of the knowledge cut-off date. Notably, this is the only non-Western LLM model in our experiment.

|  | GPT-4o | Gemini 2.0 Flash | Claude 3.7 Sonnet | DeepSeek V3 |
|---|---|---|---|---|
| Release date | 13 May 2024 | 30 January 2025 | 24 February 2024 | 26 December, 2024 |
| Knowledge cut-off date | October 2023 | August 2024 | October 2024 | unknown |

Our results show that Claude 3.7 Sonnet by Anthropic has the more recent known knowledge cut-off date, a year after GPT-4o. Gemini 2.0 Flash is also relatively updated with a knowledge cut off in August 2024, but as our results below will show, more updated digital data does not translate into divergent or updated political speech. Among almost all metrics of divergence metrics, Google's Gemini lags behind Claude 3.7 Sonnet.

There are developers working on Russian and Persian large language models, but we have not included these models due to practical reasons. GigaChat 2.0, developed by Sber, performs better in answering factual questions in Russian than DeepSeek V3 and GPT-4o. PersianMind, a recent LLM focusing on the performance with Iran's official language, Farsi, was published in January 2024, but it only achieves comparable performance with GPT-3.5-Turbo.[32] PersianLLaMA, released in December 2023, also only achieves performance comparable with older generation LLMs like GPT2.[33] Further research could be done when alternative Russian and Persian models mature.

**EXECUTION**

First, we experiment using English language in prompts and requesting English outputs. We repeat the experiment for each pairing three times and generate a synthetic dataset containing

---

[29] Bai et al., "Constitutional AI."
[30] DeepSeek-AI et al., "DeepSeek-V3 Technical Report."
[31] Vincent, "The DeepSeek Panic Reveals an AI World Ready to Blow."
[32] Rostami et al., "PersianMind: A Cross-Lingual Persian-English Large Language Model."
[33] Abbasi et al., "PersianLLaMA: Towards Building First Persian Large Language Model."

144 English commentaries. Second, we use Chinese, Russian, and Farsi (the language spoken in Iran and its neighbouring regions of Afghanistan, Iraq, Oman, Yemen, and Bahrain) to reproduce the above experiment. This helps us evaluate how different it would be for an LLM-impersonating Chinese expert speaking Chinese, as opposed to a Chinese expert speaking English. We use the API to request translating the prefix prompt into Chinese, Russian, and Farsi, and then apply the new prompts to itself. This process produces another dataset containing 108 multilingual outputs (36 outputs for each language). The total number of evaluated outputs is 252 (144 English outputs plus 108 multilingual outputs).

**SIMILARITY AND DIVERSITY METRICS**

If the outputs still converge given different prompts and different languages, we might conclude a high level of cross-cultural homogenization. The measure convergence using five similarity metrics.

We begin with TF-IDF cosine similarity, a classical lexical-based method grounded in word frequency statistics.[34] TF-IDF discounts high-frequency, low-information terms (e.g., "the," "and") and emphasizes content-bearing vocabulary.[35] When paired with cosine similarity, it allows us to quantify how similar two responses are based on their shared weighted vocabulary.[36] Recent studies indicate that cosine similarity scores above 0.6 typically denote high textual similarity.[37] This makes it well-suited for detecting wording convergence, especially in LLMs prone to formulaic repetition. TF-IDF cosine similarity has been widely applied in areas such as essay similarity evaluation,[38] novelty detection,[39] and content-based job recommendations.[40] However, it lacks semantic depth: two responses using different synonyms or paraphrases may appear unrelated, even if their meaning is near-identical.[41] As such, it is less effective when dealing with role-prompting that encourages rewording rather than reuse.

To address this limitation, we also apply SBERT cosine similarity, which evaluates semantic similarity by encoding entire sentences as contextual embeddings.[42] Unlike TF-IDF, SBERT captures meaning beyond literal word use, allowing for comparison of paraphrased content.[43] For example, "The United States seeks to contain China's influence" and "America tries to curb China's rise" may be judged dissimilar under TF-IDF but highly similar under SBERT. Prior work has shown SBERT to outperform traditional lexical metrics in tasks requiring nuanced understanding, such as semantic similarity retrieval,[44] translation closeness measurement,[45] and essay off-topic detection.[46] Although SBERT generally offers better semantic alignment,[47] its embeddings are opaque and more computationally intensive. Moreover, some research suggests that traditional TF-IDF-based methods may occasionally outperform advanced context-aware methods such as SBERT in certain structured scenarios,[48] and SBERT can overestimate similarity when two texts share topic domains but express opposing views.

---

[34] Lan, "Research on Text Similarity Measurement Hybrid Algorithm with Term Semantic Information and TF-IDF Method."
[35] Kumar et al., "Similarity Measure Approaches Applied in Text Document Clustering for Information Retrieval;" Danish et al., "Comparative Analysis of BERT and TF-IDF for Textual Semantic Similarity Assessment."
[36] Mardatillah et al., "Citation Analysis on Scientific Articles Using Cosine Similarity."
[37] Ibid.
[38] Lahitani et al., "Cosine Similarity to Determine Similarity Measure: Study Case in Online Essay Assessment."
[39] Achananuparp, Hu, and Shen, "The Evaluation of Sentence Similarity Measures."
[40] Nadar et al., "Job Recommendation by Content Filtering Using TF-IDF and Cosine Similarity."
[41] De Boom et al., "Learning Semantic Similarity for Very Short Texts;" Ortakci and Borhan, "Optimizing SBERT for Long Text Clustering."
[42] Reimers and Gurevych, "Sentence-BERT."
[43] Venkatesh Sharma et al., "Enhancing Query Relevance."
[44] Wei et al., "Index Construction and Similarity Retrieval Based on Sentence-BERT."
[45] Afzal, Rauf, and Majid, "Semantic Similarity of the Holy Quran Translations with Sentence-BERT."

While SBERT and TF-IDF operate on vector representations, we include two string-based metrics to account for exact or near-exact surface overlap. Jaccard similarity measures the proportion of shared words between two texts.[49] Its strength lies in simplicity and interpretability—especially when comparing short outputs or checking for boilerplate repetition. It is commonly employed in article searching,[50] plagiarism detection,[51] and hashing algorithms.[52] Yet Jaccard scores are highly sensitive to sentence length and word choice variation; even slight rephrasing can cause scores to drop sharply, making it less reliable for measuring meaning retention across stylistic changes.[53]

A more flexible alternative is Levenshtein similarity, which quantifies how many character-level edits are required to turn one string into another.[54] The Levenshtein similarity (or ratio) normalizes the edit distance within a 0 to 1 range.[55] It has been widely applied in contexts similar to TF-IDF, such as plagiarism detection.[56] This allows Levenshtein to detect similarity even when words have been re-ordered or slightly altered. Still, Levenshtein remains blind to semantic meaning: "peaceful rise" and "non-violent ascension" would be penalized despite expressing the same concept. It also struggles to perform effectively in automated language classification or contexts involving nuanced semantic differences.[57] Hence, it is more useful for short or formulaic texts (e.g., state media headlines) rather than rich, naturalistic language.

Finally, Word Mover's Distance (WMD) proposed by Kusner et al.[58] bridges the lexical and semantic domains by calculating the minimal cumulative "travel distance" between words in one sentence and their semantic counterparts in another, using word embeddings.[59] This makes WMD particularly powerful for detecting paraphrasing and latent meaning similarity, even when texts share no exact words. WMD has been effectively applied in areas emphasizing semantic coherence, such as interpersonal linguistic coordination,[60] essay scoring,[61] and analyzing concept engagement in texts.[62] Compared to cosine similarity, WMD is more robust in identifying latent semantic convergence, but its sensitivity also means that stylistic elaboration (e.g., using more flowery language) can inflate dissimilarity scores.

---

[46] Huang et al., "A Study of Sentence-BERT Based Essay Off-topic Detection."
[47] Danish et al., "Comparative Analysis of BERT and TF-IDF for Textual Semantic Similarity Assessment."
[48] Mandal et al., "Unsupervised Approaches for Measuring Textual Similarity between Legal Court Case Reports."
[49] Achananuparp, Hu, and Shen, "The Evaluation of Sentence Similarity Measures."
[50] Rinartha and Suryasa, "Comparative Study for Better Result on Query Suggestion."
[51] Diana and Ulfa, "Measuring Performance of N-Gram and Jaccard-Similarity Metrics."
[52] Ji et al., "Min-Max Hash for Jaccard Similarity."
[53] Wang and Dong, "Measurement of Text Similarity: A Survey."
[54] Zhang, Hu, and Bian, "Research on String Similarity Algorithm Based on Levenshtein Distance."
[55] Ibid.
[56] Li and Liu, "A Normalized Levenshtein Distance Metric."
[57] Ibid.
[58] Kusner et al., "From Word Embeddings to Document Distances."
[59] Ibid.
[60] Nasir et al., "Modeling Interpersonal Linguistic Coordination in Conversations Using Word Mover's Distance."
[61] Tashu and Horvath, "Pair-Wise: Automatic Essay Evaluation Using Word Mover's Distance."

Taken together, these five methods allow us to see different layers of resemblance in LLM outputs. Some capture repetition; others pick up shifts in tone, rephrasing, or deeper semantic alignment. We apply all five systematically across our dataset, comparing model outputs by target country, regardless of language or prompt role. These five metrics jointly enable us to capture a broad spectrum of similarity dimensions across LLM outputs, laying the groundwork for the comparative analyses presented in the next section.

**RESULTS AND INTERPRETATIONS**
Before showing all the numbers and heatmaps, we would like to highlight the most important take-aways.

In English responses, Iran politics suffers from the most homogenous output, (most frequently having the highest average scores in Levenshtein similarities, SBERT cosine similarities and the lowest average distances in Word Mover's Distance), while responses from different country perspectives about the United States politics are less homogeneous (most frequently having the lowest average scores in Jaccard similarities, SBERT cosine similarities, and the highest average distances in Word Mover's Distance). It means that the models are capable to speak about US politics from Chinese, Iranian, Russian, as well as American perspectives. The same capability of divergent political speak is not there for Iran (and to an significant extent also China).

Claude presents the most divergent English responses when prompted to impersonate experts from Iran, China, Russia, and US. We will close read Claude responses in a later section to examine the extent of actual divergence in stand points. Gemini produces the most homogeneous results across all tests in its English responses, which is an indicator of Gemini's inability to change perspectives.

DeepSeek provides an interesting contrast: it produces surprisingly diverse English language results when prompted to adopt different countries' perspectives on Iran, Russia, and US politics. The same divergence is not where in responses on Chinese politics. All languages and all perspectives present homogeneous output – a clear product of fine-tuning. This is despite the model's ability to impersonate divergent perspectives on Russia and Iran (by metrics the divergence is comparable to Claude, the top performer in divergence). Claude generates more diverse output regarding China if prompted in Chinese than it those regarding the US if prompted in English.

GPT-4o produces the most homogeneous results across all tests in its Farsi, Russian, and Chinese results. To put it more accurately, we can say the response language does not change a lot in terms of similarities for GPT-4o, while for all other models, results in the impersonated experts' own languages will produce much more divergent results. It means that no matter which language GPT-4o speaks, it gives the same responses; its monolingualism in political speak is the highest.

---
[62] Stoltz and Taylor, "Concept Mover's Distance: Measuring Concept Engagement in Texts via Word Embeddings."

**Similarity Heatmap: the hotter, the more homogeneous.**

We generate a series of heatmaps for each evaluation metric, models, and languages. Each figure below comprises 32 (4 * 8) matrices, or heatmaps, each row representing the results from each of the four models, and each column representing aggregated similarity of a target country's politics from source countries and all attempts. This information can be seen in the caption of each heatmap in the figure. For example, the top left heatmap of figure 1 is captioned "Claude 3.7 Sonnet on China," meaning that the heatmap is generated from comparing all the aggregated English outputs from the four source countries. The fifth heatmap on the top row of figure one is captioned "Claude 3.7 Sonnet (Translated) on China," meaning that the heatmap is generated from comparing all multilingual outputs (produced through prompting in Russian, Farsi, and Chinese and then translate back to English for comparison) from the four source countries. The first four columns from the left show the results with English prompts and the remaining columns show the results with prompts in the language used by the perspective "expert" role (the comparisons are carried out among the responses translated back in English).

For each heatmap inside a figure, a cell represents the average value of the similarity scores between the perspective expert role specified in the x- and y-axis (hence the comparisons between the same perspective expert role on the same country are 1 (0 for Word Mover's Distance) as two responses in comparison are identical. The experiment was carried out repeatedly three times, and the average values were calculated by aggregating the results of each attempt and averaging them, i.e. the sum of the similarity scores of each comparison in the three attempts divided by three.

The results from all five metrics are positively correlated. If Claude scores less than Gemini by one metric, which indicates Claude is capable of more divergent results than Gemini by that metric, Claude also tend to score similarly low on other metrics. The five metrics are relatively consistent in its evaluation of results based on models as well as on target countries.

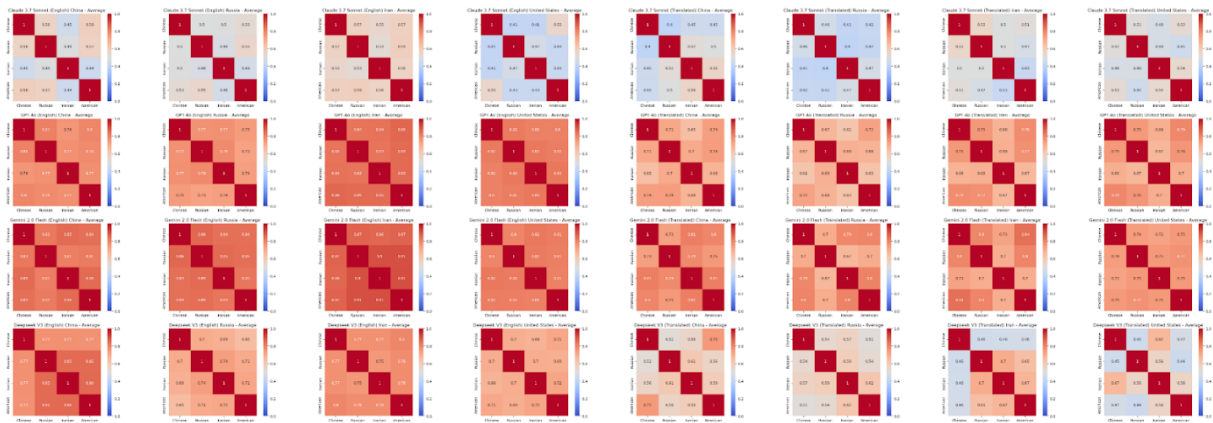

Figure 1. TF-IDF Cosine similarity matrices across different models.

Comparing the relative performances of the four AI products with TF-IDF cosine similarity (Figure 1), we could first observe that LLM tends to provide more similar responses when prompted with the same language (English) despite different country roles. Across different AI products, Anthropic's Claude 3.7 Sonnet provides responses with the lowest similarities when compared with the responses on the same country from a different role, while Google's Gemini 2.0 Flash, OpenAI GPT-4o, and Deepseek V3 offer significantly more similar responses from different perspective roles.

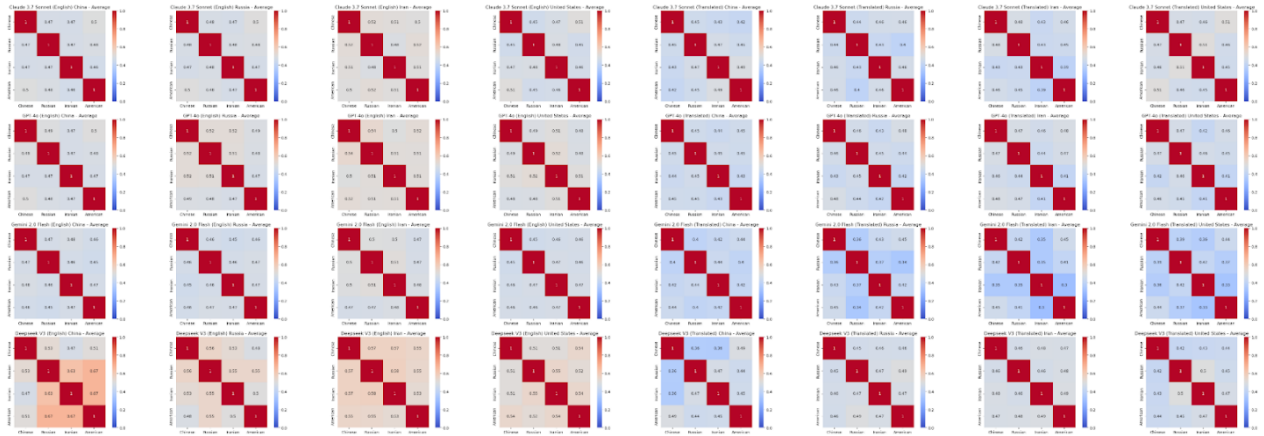

Figure 2. Levenshtein similarity matrices across different models.

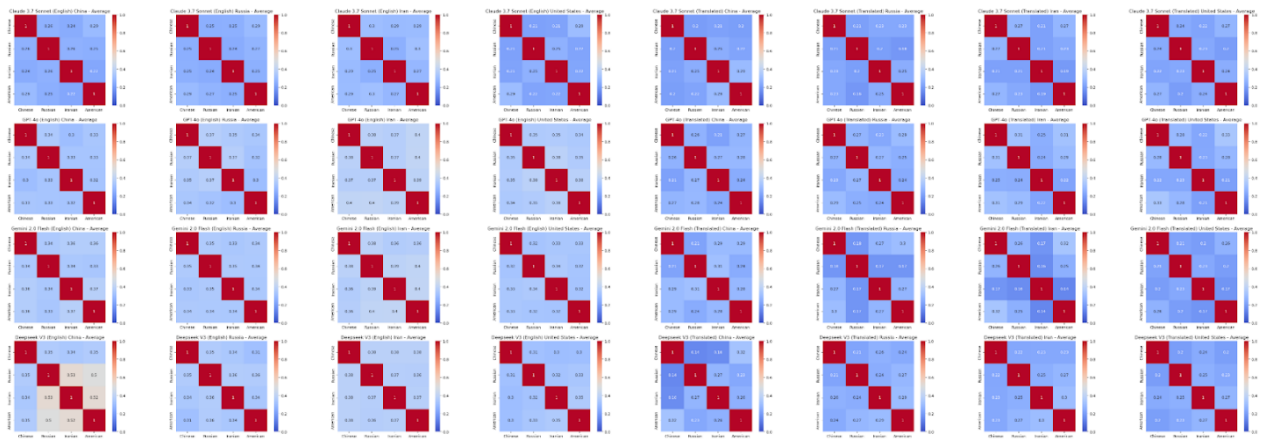

Figure 3. Jaccard similarity matrices across different models.

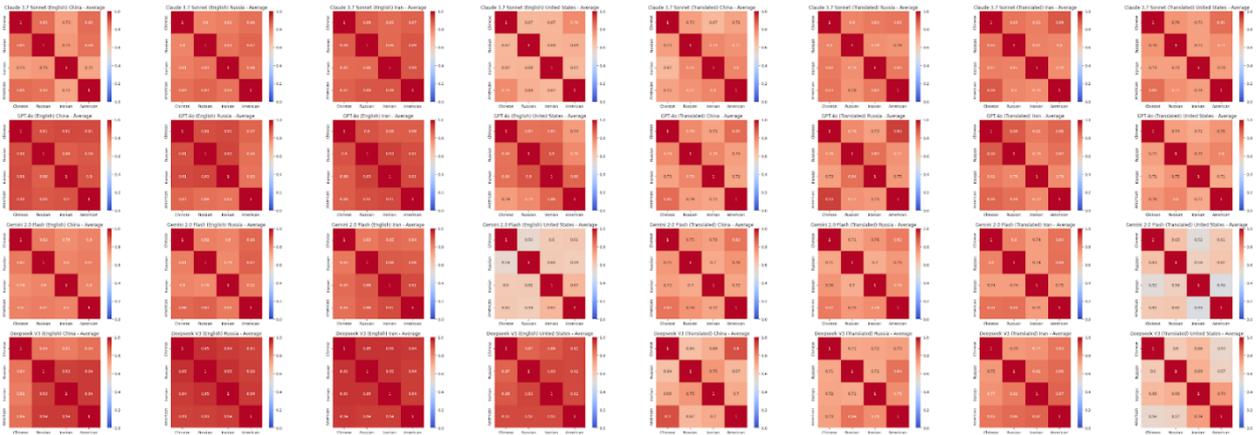

Figure 4. SBERT Cosine similarity matrices across different models.

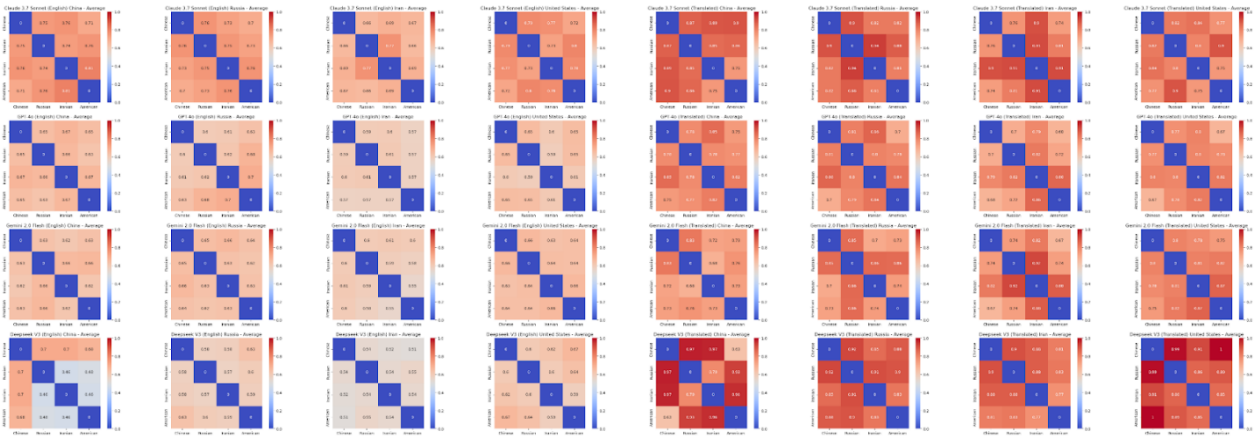

Figure 5. Word Mover's Distance matrices across different models.

**More Detailed Reading of One Metric**

To assess the consistency and divergence of LLM-generated political discourse, we focus on the representative metric TF-IDF cosine similarity. Although our broader analysis incorporates five distinct similarity measures, we have selected TF-IDF for closer examination because of its clarity and effectiveness in capturing patterns of lexical repetition. The patterns observed through TF-IDF are broadly consistent with those identified by the other metrics—SBERT, Jaccard, Levenshtein, and Word Mover's Distance (see Figures 2 to 5). For instance, we have generated the following texts to show the consistence of these metrics at a more intuitive level. The following texts are two AI-generated responses on cultural policies in the United States. Two responses are required to be "similar" when prompting the large language model, and they are similar in terms of theme, content, word usage, and examples.

| Response A | Response B |
|---|---|
| U.S. cultural policies promote diversity and artistic expression through initiatives like the NEA (National Endowment for the Arts), support for public broadcasting, preservation of historic sites, and funding for cultural | U.S. cultural policies support artistic innovation and diversity through programs like the NEA, historic site preservation, public broadcasting, and cultural festivals. These initiatives aim to promote inclusion, protect |

festivals. These policies aim to enrich national identity, foster creativity, and preserve cultural heritage.

heritage, and encourage creativity across communities, strengthening national identity and fostering a vibrant, multicultural society.

Similarity scores for responses A and B and its explanation

| Cosine Similarity | | Levenshtein Ratio | Jaccard Similarity | Word Mover' Distance |
|---|---|---|---|---|
| TF-IDF | SBERT | | | |
| 0.5684 | 0.9300 | 0.5677 | 0.3594 | 0.0906 |
| Both metrics are sensitive to semantic differences. SBERT scores remains more stable when outputs are different in length. | | Levenshtein Ratio would increase significantly when we swap out the synonyms with exact phrasing. | Jaccard Similarity is less sensitive to semantic differences. | The smaller the number, the less distance, the more similar. |

The TF-IDF cosine similarity heatmaps (see above Figure 1) display a series of matrices, each corresponding to a specific model and target country. These matrices represent the average pairwise similarity between outputs generated by different expert roles. In each matrix, the first four columns correspond to prompts issued in English, while the remaining four reflect prompts in the native language associated with the respective expert role. The diagonal values indicate perfect self-similarity, while the off-diagonal values capture the degree of variation among different role-based responses to the same country-specific prompt.

Several recurring trends emerge from the heatmaps. Across all four models, English prompts tend to produce more similar outputs than prompts given in native languages. This trend suggests that prompting in English may encourage more uniform phrasing and structure, potentially reflecting the underlying training data bias toward English-language sources. Gemini 2.0 Flash is a particularly notable example: when asked about Iranian politics in English, it produces highly homogeneous outputs, with nearly all off-diagonal similarity scores exceeding 0.95. This pattern implies that the model generates nearly identical responses across different roles, even for politically sensitive topics.

In contrast, Claude 3.5 Sonnet demonstrates a higher degree of variation. When prompted in Chinese to describe China, the similarity scores between expert roles frequently drop below 0.80. This indicates that Claude is more responsive to linguistic shifts in the prompt, leading to more differentiated outputs across roles. GPT-4o, by comparison, exhibits only minimal differences between English and native-language prompts. The similarity scores remain relatively high in both settings, suggesting that the model's translation pipeline tends to produce outputs with fixed phrasing regardless of the input language. DeepSeek V3 presents the sharpest contrast between English and native-language prompting. In several cases, such as responses about

Russia and Iran, native-language prompts result in significantly lower similarity scores, indicating greater stylistic or lexical diversity.

Taken together, these findings point to a consistent pattern: the language of prompting appears to have a greater impact on output variation than role assignment. The heatmaps help identify which models are more responsive to shifts in input language and which tend to generate more uniform political discourse across roles. Building on these aggregated observations, the next section turns to specific examples that illustrate these trends in detail by comparing pairs of the most and least similar outputs.

**Close-Readings:**
**ChatGPT-4o outputs on the politics of Iran from Russian and US perspectives in English**
Cosine Similarity: 0.88
In real-life situations, we would expect the Russian perspective to differ from the US perspective regarding Iran's politics, even if both roles speak in English. Yet in the output, we see almost exact wording, formatting, paragraphing, and length from the Russian and US perspectives (see Table 1).

*Table 1. The Most Similar Results, cosine similarity 0.88.*

| Russian Perspective, in English | US Perspective, in English |
|---|---|
| Certainly. Iran's political structure is unique and combines elements of a theocracy with republican principles. The country was transformed into an Islamic Republic after the 1979 Iranian Revolution, which overthrew the Pahlavi monarchy. Here's a summary of its political landscape: | Certainly! Iran's political system is a unique blend of theocratic and democratic elements, structured around the principles of the Islamic Republic, as established after the 1979 Iranian Revolution. Here is an overview of the key components of Iran's political system: |
| 1. **Supreme Leader**: At the apex of Iran's political hierarchy is the Supreme Leader, who holds considerable influence over all branches of government and the military. The Supreme Leader is regarded as the highest religious and political authority in Iran, and their duties involve overseeing the judiciary, state broadcasting, and the armed forces. Ali Khamenei has held this position since 1989. | 1. **Supreme Leader**: The Supreme Leader holds the highest authority in Iran and has significant influence over all branches of government, including the military and media. This role was first held by Ayatollah Ruhollah Khomeini and is currently held by Ayatollah Ali Khamenei. The Supreme Leader is chosen by the Assembly of Experts and can influence policies, appointments, and the country's direction. |
| 2. **President**: The President of Iran is the highest-ranking official elected by the public. While the role is significant, it is subordinate to the Supreme Leader. The President is responsible for the implementation of the Constitution, economic policy, and foreign affairs. Ebrahim Raisi has been serving as the President since August 2021. | 2. **President**: The President of Iran is the second-highest official and is elected by popular vote for a four-year term, with a maximum of two consecutive terms. The President manages the executive branch and implements laws, but their power is limited by the Supreme Leader. The President also represents Iran at international forums. As of my last update, Ebrahim Raisi holds this position. |
| 3. **Guardian Council**: This body consists of twelve members, six clerics appointed by the Supreme Leader and six jurists nominated by the judiciary and approved by the Parliament. The Guardian Council vets candidates for elected office and can veto legislation passed by the Parliament if it is deemed inconsistent with the Constitution or Islamic law. | 3. **Parliament (Majlis)**: The Islamic Consultative Assembly, or Majlis, is Iran's legislative body, consisting of 290 members elected every four years. It drafts legislation, approves budgets, and can impeach ministers. However, legislation must be approved by the Guardian Council and align with Islamic law. |
| 4. **Parliament (Majlis)**: Iran's unicameral legislative body has 290 members, elected every four years. The Parliament drafts legislation, ratifies international treaties, and approves the national budget. | 4. **Guardian Council**: This 12-member body is responsible for vetting candidates for elections and ensuring that legislation aligns with the Constitution and Islamic law. Six members are appointed by the Supreme Leader, and six are nominated by the judiciary and approved by the Majlis. The Council plays a critical role in shaping the political landscape by influencing who can run for office. |
| 5. **Expediency Discernment Council**: This council mediates disputes between the Parliament and the Guardian Council and advises the Supreme Leader. It is composed of senior officials and appointed members. | 5. **Assembly of Experts**: An 88-member body elected by the people for eight-year terms, it is responsible for appointing and overseeing the Supreme Leader. While its theoretical authority is significant, in practice, it often acts in line with the broader political structure. |
| 6. **Assembly of Experts**: An 88-member body of senior clerics tasked with electing the Supreme Leader and monitoring their performance. Its members are directly elected by the people, but candidates must be approved by the Guardian Council. | 6. **Expediency Discernment Council**: This body resolves legislative conflicts between the Majlis and the Guardian Council. It also serves as an advisory body to the Supreme Leader. |
| 7. **Revolutionary Guard**: The Islamic Revolutionary Guard Corps (IRGxC) plays a significant role in Iran's politics. It is an elite military force established after the revolution to protect the Islamic system and prevent foreign intervention. The IRGC also holds substantial economic power through its various enterprises. | 7. **Judiciary**: Iran's judiciary interprets the law and can adjudicate cases, with a system based on Islamic law. The head of the judiciary is appointed by the Supreme Leader. |
| 8. **Political Climate and Challenges**: Iran's politics are often marked by tensions between hardliners and reformists. Hardliners favor strict adherence to religious principles, while reformists seek more progressive changes and openness, particularly in international relations and media. Internationally, Iran faces challenges related to its nuclear program, sanctions, human rights concerns, and geopolitical influence in the region. | Iranian politics are marked by tensions between conservative and reformist factions. Conservatives generally emphasize strict adherence to Islamic principles and the goals of the 1979 Revolution, while reformists advocate for more political freedoms and engagement with the international community. This dynamic creates a complex and often contentious political environment. |
| 9. **Societal Impact**: Domestic politics in Iran significantly impact civil liberties, media freedom, and women's rights. Periodic protests and movements occur, often triggered by economic conditions, political repression, and violations of personal freedoms. | Iran's international relations—especially its nuclear program and regional interventions—are significant topics in its foreign policy. The country faces international sanctions, primarily from the United States, which impact its economy and global relations. Issues such as human rights, press freedom, and political activism are also challenging elements within Iran's political landscape. |
| Overall, Iran's political environment is complex and heavily influenced by its Islamic ideology, with significant power held by unelected bodies under the supervision of the Supreme Leader. | |

The two expert texts share common lexicon, such as "human rights," "civil liberties," and "(media) freedom," found in the Western mainstream media reporting about Iran. Context-specific terms such as "Supereme Leader," "President," "Guardian Council," "Parliament (Majlis)," "Expediency Discernment Council," and "Assembly of Experts" and their descriptions are nearly identical in both accounts. Both frame the narratives in terms of "international/social impacts," avoiding deeper engagement with their implications. The only difference is that the Russian expert frames Iran's political turmoil as a struggle between "hardliners" and

"reformists," and the difference does not reflect the two countries' divergent diplomatic approach to Iran.

**Gemini on the politics of Iran from the Russian perspective in Russian and the US perspective in English**

Cosine Similarity: 0.87

To show the degree of Gemini's homogenization, we chose to closely read the results from Gemini's Russian expert speaking in Russian and US expert speaking in English, both on Iran's politics. Language variation introduces significant perspectival shifts in Claude, but not for Gemini. Except for a difference in tone (US output being more verbose, thus longer and slightly more casual), the content remains roughly the same. The outputs register no political, social, or cultural differences between the two perspectives (see Table 2) .

*Table 2. Similar Results, Cosine Simiarity at 0.87*

| Russian Perspective, in Russian, then translated into English | US Perspective, in English |
|---|---|
| **A brief overview of Iran's political system:** Iran is an **Islamic republic**, and its political system is unique. It combines elements of theocracy and democracy, but with an emphasis on religious legitimacy. Key features: * **Supreme Leader (Rahbar):** The highest authority in Iran. Currently, it is Ayatollah Ali Khamenei. He has enormous powers, including defining the overall policy of the country, commanding the armed forces, approving candidates for presidential elections, appointing heads of the judiciary and other important positions. In fact, he is the main arbiter in all political matters. * **President:** Elected by popular vote for a four-year term (with the possibility of re-election once). The President is the head of the executive branch and is responsible for the day-to-day management of the country. However, his powers are limited by the decisions of the Supreme Leader. The current president is Ebrahim Raisi. * **Majlis (Islamic Consultative Assembly):** A unicameral parliament, elected by popular vote. The Majlis develops and adopts laws, approves the state budget and oversees the activities of the government. However, all bills must be approved by the Guardian Council of the Constitution. * **Guardian Council of the Constitution:** Consists of six theologians appointed by the Supreme Leader and six lawyers elected by the Majlis from a list proposed by the Chief Justice. The Guardian Council monitors the compliance of bills and government actions with the Constitution and Sharia (Islamic law). It also approves candidates for presidential elections and in the Majlis, which allows it to influence the political landscape of the country. * **Assembly of Experts:** Elected by popular vote and consists of religious figures. Its main function is to elect the Supreme Leader and monitor his activities. **Main political forces and factions:** In Iranian politics, there are no formal political parties in the Western sense. Instead, various political factions and groupings operate, which can be conditionally divided into the following: * **Conservatives (Principalists):** Support the preservation of Islamic values, a strong role of the state in the economy and a tough line in foreign policy. * **Reformists:** Advocate for the expansion of political freedoms, the development of a market economy and the improvement of relations with the West. * **Moderates:** Occupy an intermediate position between conservatives and reformists, striving for a pragmatic solution to problems. It is important to note that these factions are not monolithic, and there are various currents within them. **Main political problems and challenges:** Iran faces a number of political problems and challenges, including: * **Economic difficulties:** High levels of unemployment, inflation and corruption, exacerbated by international sanctions. * **Social discontent:** Growing public discontent with the economic situation, political restrictions and social injustice. * **Foreign policy challenges:** Strained relations with the United States, Israel, and a number of other countries, as well as participation in regional conflicts. * **The issue of the nuclear program:** Iran's nuclear program is the subject of intense international controversy and has a significant impact on the country's foreign policy. * **Human rights:** The human rights situation in Iran raises serious concerns among international organizations. **Iran's foreign policy:** Iran's foreign policy is determined by its religious beliefs, its desire for regional leadership, and its opposition to the United States and Israel. Iran actively supports various Shiite groups in the region and interferes in politics in conflicts in Syria, Iraq, Lebanon and Yemen. It also develops relations with Russia, China, and other countries that are not part of the Western bloc. **In conclusion:** Iran's political system is complex and unique. It represents a combination of theocratic and democratic elements, in which religious leaders play a key role. The country faces a number of serious political, economic and social problems, and its foreign policy is characterized by tense relations with the West and active participation in regional conflicts. | Okay, let's delve into the fascinating and complex world of Iranian politics. It's a system quite unlike anything else in the world, blending elements of theocracy and democracy in a unique and often volatile manner. **Key Characteristics of the Iranian Political System:** * **The Dual Structure: Theocratic and Republican Elements:** Iran's political system is often described as a "theocracy disguised as a republic." This means that while there are elections and a parliament (the Majlis), ultimate authority resides with religious figures, particularly the Supreme Leader. * **The Supreme Leader (Rahbar):** Currently Ayatollah Ali Khamenei, the Supreme Leader is the most powerful figure in Iran. He is the head of state and commander-in-chief of the armed forces. He has the final say on all major policy decisions, including foreign policy, national security, and economic matters. He is selected by the Assembly of Experts. * **The President (Rais-Jumhur):** Currently Ebrahim Raisi, the President is the head of the executive branch and is responsible for the day-to-day administration of the country. He is elected by popular vote, but candidates must be vetted by the Guardian Council to ensure their loyalty to the Islamic Republic. * **The Parliament (Majlis):** The Majlis is a unicameral legislature elected by popular vote. It has the power to legislate, approve the budget, and ratify treaties. However, all laws passed by the Majlis must be approved by the Guardian Council. * **The Guardian Council (Showra-ye Negahban):** This 12-member body is composed of six clerics appointed by the Supreme Leader and six jurists nominated by the judiciary and approved by the Majlis. The Guardian Council has the power to veto legislation passed by the Majlis and to vet candidates for elections. * **The Assembly of Experts (Majles-e Khobregan):** This body is composed of elected clerics and is responsible for selecting the Supreme Leader and monitoring his performance. * **The Role of Clerics:** Clerics play a central role in Iranian politics. They hold many key positions in the government, the judiciary, and the armed forces. They also exert considerable influence over public opinion through their control of mosques and religious institutions. * **Factions and Political Alignments:** Iranian politics is often characterized by factionalism. There are generally two main political factions: * **Principalists/Conservatives:** Generally, they are loyal to the principles of the Islamic Revolution, favor a strong role for the state in the economy, and advocate a hardline foreign policy. They are often wary of Western influence and prioritize Iran's national security and regional interests. * **Reformists:** They tend to favor greater political and social freedoms, economic liberalization, and improved relations with the West. They are often critical of the hardline policies of the conservatives and advocate for reforms within the existing system. *Important Note:* These factions are often fluid and overlapping, and individuals can shift their allegiances over time. It's also crucial to avoid simplistic labels. The landscape is far more nuanced than "hardliner" versus "moderate." * **Foreign Policy:** Iran's foreign policy is driven by a combination of factors, including its revolutionary ideology, its national interests, and its regional ambitions. Key aspects include: * **Anti-Americanism and Anti-Zionism:** Since the 1979 revolution, Iran has maintained a strong anti-American and anti-Zionist stance. This is rooted in its opposition to US foreign policy in the Middle East and its support for the Palestinian cause. * **Regional Influence:** Iran seeks to exert influence in the Middle East, particularly in countries like Iraq, Syria, Lebanon, and Yemen. It supports various Shia militias and political groups in these countries. * **Nuclear Program:** Iran's nuclear program has been a major source of international concern. Iran maintains that its nuclear program is for peaceful purposes, but Western powers fear that it could be used to develop nuclear weapons. * **Relationship with Russia and China:** In recent years, Iran has strengthened its ties with Russia and China, seeking to counter Western pressure and expand its economic and political partnerships. **Challenges and Issues Facing Iranian Politics:** * **Economic Problems:** Iran's economy has been struggling for years, due to a combination of factors, including international sanctions, mismanagement, and corruption. High unemployment, inflation, and inequality are major sources of discontent. * **Social Dissatisfaction:** There is widespread social dissatisfaction in Iran, particularly among young people and women. Many Iranians are calling for greater political and social freedoms, as evidenced by recent protests. * **Human Rights:** Iran's human rights record is a major concern. The government has been criticized for its suppression of dissent, its use of the death penalty, and its discrimination against women and minorities. * **Succession:** The eventual succession of the Supreme Leader is a significant issue. The Assembly of Experts will have to choose a new leader, and the process could be contentious and could potentially lead to political instability. * **International Relations:** Iran's relations with the West, particularly with the United States, remain strained. The nuclear issue, Iran's support for regional proxies, and its human rights record are all major sources of tension. **Key Considerations for Understanding Iranian Politics:** * **Historical Context:** Understanding Iran's history, particularly the 1979 revolution, is crucial for understanding its current political system and its foreign policy. * **Religious Ideology:** Shi'a Islam is a central component of Iran's political identity and informs many of its policies. * **Power Dynamics:** The complex interplay of power between the Supreme Leader, the President, the Parliament, the Guardian Council, and other institutions must be understood. * **Domestic Concerns:** Public opinion, economic issues, and social grievances play a significant role in shaping Iranian politics. **In conclusion, Iranian politics is a complex and dynamic field. It is shaped by a unique blend of theocracy and republicanism, by factionalism and by the legacy of the 1979 revolution. Understanding the key actors, institutions, and issues is essential for analyzing Iran's domestic and foreign policy.** What specific aspects of Iranian politics are you most interested in? Knowing that will allow me to provide even more focused and helpful information. For example, are you interested in: * The current state of the nuclear negotiations * The role of the Revolutionary Guard (IRGC) in Iranian politics and the economy? * The prospects for political reform in Iran? * The impact of sanctions on Iranian society? * The dynamics between different political factions? Let me know! |

The two Gemini experts group the lexicon like "human rights," "injustice/inequality," and "political restrictions/suppression" with Iranian politics. The grouping is prevalent in the Western official reports on Iran's political surveillance and critiques of its media manipulation, for example, found in the United Nations[63] and BBC NEWS[64]. In addition to the overlapping interpretations of Iran's key political characteristics, the Russian and the US experts speak almost identically about Iranian positionality in international relationships, emphasizing the country as either "not part of the Western bloc" or "counter(ing) Western pressure." Both hint at Iran's threat of its nuclear program, reflecting the latent fear in the West (US, especially)[65]. Both conclude Iranian politics in exact one voice on its complexity and religious-intervention uniqueness.[66] Just one intuitive difference lies in the Iranian political fractions, which the Russian expert frames as Conservatives/Reformists/Moderates.

---

[63] United Nations Human Rights Office, "Iran Government Continues Systematic Repression and Escalates Surveillance."
[64] BBC News, "Iran Media Guide."
[65] Ferragamo et al., "What Are Iran's Nuclear and Missile Capabilities?"

**GPT-4o on the politics of Iran from the Iranian perspective in Farsi and the US perspective in English**

Cosine Similarity: 0.55

We begin to see the real difference in tone, content, and approaches to the prompts from GPT-4o's Iranian expert's Farsi results on Iran and the US expert's English results on Iran – although the divergence is not as much as we had expected (0.55 is still quite a high similarity score). The two outputs register different talk points. The Farsi output is significantly shorter than the English output.

*Table 3. Moderate Difference, Cosine Similarity at 0.55*

| Iranian Perspective, in Farsi, then translated into English | US Perspective, in English |
|---|---|
| Yes, certainly. Iran is a complex and multi-layered country in terms of international politics. Politics in Iran is influenced by a set of internal and external factors, some of which include government policies, regional and international interactions, and economic and social issues.<br><br>Domestically, Iran is an Islamic Republic led by the Supreme Leader. The leadership in Iran has significant power in major national decision-making and sets general policies. The President, elected by the people, is the head of the executive branch and is responsible for implementing the country's policies.<br><br>Internationally, Iran's foreign policy is influenced by various issues such as the country's nuclear program, economic sanctions, and relations with neighboring countries and global powers. As a major player in the Middle East, Iran plays an important role in regional issues like the crises in Syria, Yemen, and Iraq.<br><br>Issues like the Iran nuclear deal (JCPOA) and relations with Western powers, especially the United States, have been among the most important topics in Iran's foreign policy in recent years. Interactions with other countries such as Russia, China, and India are also of special importance to Iran.<br><br>In addition to these, economic issues, influenced by sanctions and internal policies, also play a crucial role in shaping Iran's policies. Managing oil and gas resources and efforts to expand trade relations with other countries are key topics in Iran's economic policymaking.<br><br>If you have any specific questions or need more details on a particular subject, please let me know. | Certainly! Iran's political system is a unique blend of theocratic and democratic elements, structured around the principles of the Islamic Republic, as established after the 1979 Iranian Revolution. Here is an overview of the key components of Iran's political system:<br>1. **Supreme Leader**: The Supreme Leader holds the highest authority in Iran and has significant influence over all branches of government, including the military and media. This role was first held by Ayatollah Ruhollah Khomeini and is currently held by Ayatollah Ali Khamenei. The Supreme Leader is chosen by the Assembly of Experts and can influence policies, appointments, and the country's direction.<br>2. **President**: The President of Iran is the second-highest official and is elected by popular vote for a four-year term, with a maximum of two consecutive terms. The President manages the executive branch and implements laws, but their power is limited by the Supreme Leader. The President also represents Iran at international forums. As of my last update, Ebrahim Raisi holds this position.<br>3. **Parliament (Majlis)**: The Islamic Consultative Assembly, or Majlis, is Iran's legislative body, consisting of 290 members elected every four years. It drafts legislation, approves budgets, and can impeach ministers. However, legislation must be approved by the Guardian Council and align with Islamic law.<br>4. **Guardian Council**: This 12-member body is responsible for vetting candidates for elections and ensuring that legislation aligns with the Constitution and Islamic law. Six members are appointed by the Supreme Leader, and six are nominated by the judiciary and approved by the Majlis. The Council plays a critical role in shaping the political landscape by influencing who can run for office.<br>5. **Assembly of Experts**: An 88-member body elected by the people for eight-year terms, it is responsible for appointing and overseeing the Supreme Leader. While its theoretical authority is significant, in practice, it often acts in line with the broader political structure.<br>6. **Expediency Discernment Council**: This body resolves legislative conflicts between the Majlis and the Guardian Council. It also serves as an advisory body to the Supreme Leader.<br>7. **Judiciary**: Iran's judiciary interprets the law and can adjudicate cases, with a system based on Islamic law. The head of the judiciary is appointed by the Supreme Leader.<br>Iranian politics are marked by tensions between conservative and reformist factions. Conservatives generally emphasize strict adherence to Islamic principles and the goals of the 1979 Revolution, while reformists advocate for more political freedoms and engagement with the international community. This dynamic creates a complex and often contentious political environment.<br>Iran's international relations—especially its nuclear program and regional interventions—are significant topics in its foreign policy. The country faces international sanctions, primarily from the United States, which impact its economy and global relations. Issues such as human rights, press freedom, and political activism are also challenging elements within Iran's political landscape. |

The two experts here provide nearly the reverse insights into the Iranian political structure. Despite the similarity in describing the functionality of the Supreme Leader and the President, the Iranian expert says nothing about the country's whole domestic political structure. Besides, the Iranian expert, to a large extent, peels off the lexicons, like "human rights" and "press freedom," which are embedded in the politically accusatory accounts by the US experts. Although both experts adopt a measured tone in telling the country's international politics, including the nuclear program, the Iranian perspective is much more "multilayered," mentioning the relations with the Middle East, Russia, and China, in addition to Western powers like the US.

**Claude on the politics of China from the Iranian perspective in Farsi and the US perspective in English**

Cosine Similarity: 0.37

The real difference comes in the pair regarding Chinese politics from Iran and the US, which currently have contrasting diplomatic approaches to China. The two outputs are similar in length, which indicates a relatively equitable treatment of both Farsi and English (contrasting ChatGPT's output above).

---

[66] CAO, "Monolingual LLMs in the Age of Multilingual Chatbots."

*Table 4. Extreme Difference, Cosine Similarity at 0.37*

| Iranian Perspective, in Farsi, then translated into English | US Perspective, in English |
|---|---|
| ## Analyzing Chinese Politics: An Iranian International Relations Perspective<br><br>As an Iranian expert in international politics, I can offer some observations about China's political system and its global positioning.<br><br>China operates under a single-party system led by the Chinese Communist Party (CCP), which maintains tight control over governmental affairs, media, and social discourse. Under President Xi Jinping, we've witnessed significant centralization of power, including the removal of presidential term limits in 2018, which allows him to remain in office indefinitely.<br><br>From an Iranian perspective, several aspects of Chinese politics are particularly noteworthy:<br><br>**Economic pragmatism with political control**: China has successfully implemented market reforms while maintaining strict political control - creating a model distinct from Western liberal democracy that many developing nations study.<br><br>**Belt and Road Initiative (BRI)**: This massive infrastructure project represents China's effort to expand influence across Asia, Africa, and parts of Europe. For Iran, this has meant increased Chinese investment despite international sanctions.<br><br>**US-China competition**: The growing rivalry between these powers creates both opportunities and challenges for countries like Iran that seek to navigate between competing global blocs.<br><br>**China's non-interference principle**: Unlike Western powers, China officially maintains a policy of non-intervention in other countries' internal affairs, though this principle is increasingly tested as China's global interests expand.<br><br>The Chinese system demonstrates that economic development need not follow Western political models, which is significant for many countries seeking their own development paths within their cultural and historical contexts. | # Politics in China: An Overview<br><br>As an expert in international politics, I can provide you with an analysis of China's political system and its key characteristics.<br>## Political Structure<br>China operates as a one-party socialist state under the exclusive leadership of the Chinese Communist Party (CCP). Unlike Western democracies with competitive multi-party systems, China's political framework is built around the CCP's central role in governance and policymaking.<br>## Key Institutions<br>- **National People's Congress (NPC)**: Technically the highest organ of state power, though largely functions to formalize CCP decisions<br>- **State Council**: The chief administrative authority, headed by the Premier<br>- **Central Military Commission**: Controls the armed forces<br>- **Chinese Communist Party**: The paramount political institution with approximately 95 million members<br>## Leadership<br>The current leadership is headed by Xi Jinping, who serves in three critical positions:<br>- General Secretary of the CCP<br>- President of the People's Republic of China<br>- Chairman of the Central Military Commission<br>Xi's tenure has been marked by significant centralization of power, especially following the 2018 constitutional changes that removed presidential term limits.<br>## Governance Characteristics<br>- Strong emphasis on stability and social order<br>- Centralized decision-making<br>- Limited political pluralism<br>- Extensive state involvement in economy and society<br>- Focus on economic development as a source of legitimacy<br>## Current Trends<br>- Increasing emphasis on national rejuvenation and the "China Dream"<br>- Greater assertiveness in foreign policy<br>- Enhanced technological controls and surveillance capabilities<br>- Continued economic reform within the context of state guidance<br><br>Would you like me to elaborate on any particular aspect of China's political system? |

The Iranian and the US experts impersonated by Claude produce the most divergent results on China's politics. The two outputs both cover China's international contributions, but in almost opposite directions. The Iran expert stems from the implications for its domestic future and economic benefits and correspondingly sings high for China's "non-interference principle" and Belt and Road Initiative (BRI). The US expert concludes that China's diplomatic approach is "greater assertiveness in foreign policy," framed as current trends following "the national rejuvenation and the 'China Dream'." We see distinction in words, especially adjectives and adverbs. The Iranian expert borrows phrases such as "particularly noteworthy" and "successfully implemented" to reflect the positive sentiments towards China. When speaking of the specific political structure besides the President's unlimited centralization of power, the US expert demonstrates proficiency in detailing China's profile, reflecting the process of "othering" in this Western-based large language model. The Farsi-prompted output excludes the corresponding content in its Iranian narrative.

We can continue to read the rest of the 256 results from this dataset and see the same ideolexics largely spread across outputs concerning Iran and Russia. DeepSeek's China politics responses – the most homogeneous across all metrics and all models – are fine-tuned to reflect the expected uniformity of the Chinese internet.

**CONCLUSION**

This study has focused on presenting the empirical evidence of LLMs' flattening of diversity in political commentary. It has not engaged with politics and literature as traditional disciplinary approaches would; instead, it has taken cues from critical theory and speculates on the future trajectory of politics discourse in an era increasingly shaped by Large Language Model (LLM)-driven agentic systems.

The results from this study show that despite the volatile, dynamic political realities in these regions (exemplified by recent Middle Eastern conflicts that leads to changes daily or even hourly), LLMs pretrained predominantly on Anglophone web, fine-tuned with less-resource language database, and prompted with role-assignment for specific agentic tasks, still reproduce static, uniform characterizations of places often framed as "enemies of liberal democracy" by global mainstream media.

The core of this paper is empirical and quantitative. The goal is to show how empirical studies done with layman-accessible computational methods and synthetic data generated by LLMs could be aligned with critical-theoretical discussions. This experiment is meant to produce empirical evidence that shows the extent to which LLM-generated discourse can amplify ideological homogeneity, particularly toward states frequently framed by global mainstream media as adversaries of liberal democracy. The experiment is designed in a way that is replicable by a layman so that more humanists can scrutinize how LLMs may entrench reductive representations and how we can improve them.